\documentclass{emulateapj}
\usepackage{amssymb}
\usepackage{amsmath}

\newcommand{\mbh}{\ensuremath{M_{\rm{BH}}}\,}

\newcommand{\rev}[1]{#1}
\newcommand{\revt}[1]{ #1}

\received{October 20, 2017}
\revised{November 28, 2017}
\accepted{January 6, 2018}
\shorttitle{Balmer line absorption in two luminous LoBAL quasars at $z\sim1.5$}
\shortauthors{Schulze et al.}
\begin{document}

\title{Discovery of strong Balmer line absorption in two luminous LoBAL quasars at $z\sim1.5$}

\author{Andreas Schulze\altaffilmark{1,6}, Toru Misawa\altaffilmark{2}, Wenwen Zuo\altaffilmark{3}, Xue-Bing Wu\altaffilmark{4,5}}
\email{E-mail: andreas.schulze@nao.ac.jp}

\altaffiltext{1}{National Astronomical Observatory of Japan, Mitaka, Tokyo 181-8588, Japan}
\altaffiltext{2}{School of General Education, Shinshu University, 3-1-1 Asahi, Matsumoto, Nagano 390-8621, Japan}
\altaffiltext{3}{Shanghai Astronomical Observatory, Shanghai 200030, China}
\altaffiltext{4}{Department of Astronomy, Peking University, Yi He Yuan Lu 5, Hai Dian District, Beijing 100871, China}
\altaffiltext{5}{Kavli Institute for Astronomy and Astrophysics, Peking University, Beijing 100871, China}
\altaffiltext{6}{EACOA Fellow}



\begin{abstract}
We present the discovery of strong Balmer line absorption in H$\alpha$ to H$\gamma$ in two luminous low-ionization broad absorption line quasars (LoBAL QSOs) at $z\sim1.5$, with black hole masses around $10^{10}\ M_\odot$ from near-IR spectroscopy. There are only two previously known quasars at $\rev{z>1.0}$ showing Balmer line absorption.
SDSS~J1019+0225 shows blueshifted absorption by $\sim1400$~km s$^{-1}$ with an H$\alpha$ rest-frame equivalent width of 13~\AA{}. In SDSS~J0859+4239 we find redshifted absorption by $\sim500$~km s$^{-1}$ with an H$\alpha$ rest-frame equivalent width of 7~\AA{}. The redshifted absorption could indicate an inflow of high density gas onto the black hole, \revt{though} we cannot rule out alternative interpretations.
The Balmer line absorption in both objects appears to be saturated, indicating partial coverage of the background source by the absorber. We estimate the covering fractions and optical depth of the absorber and  derive neutral hydrogen column densities, $\rev{N_{\rm{H\,I}}} \sim 1.3\times 10^{18}$~cm$^{-2}$ for SDSS~J1019+0225 and $\rev{N_{\rm{H\,I}}}\sim9\times 10^{17}$~cm$^{-2}$  for SDSS~J0859+4239\revt{.}  In addition, the optical spectra reveal also absorption troughs in \ion{He}{1}$^*$ $\lambda3889$ and $\lambda3189$ in both objects.
\end{abstract}




\section{Introduction}
Broad Absorption Line quasars (BAL QSOs) represent an important sub-population of the AGN phenomenon. They show absorption lines in their rest-frame UV-lines with velocity widths $>2000$~km~s$^{-1}$ which are usually blueshifted  with velocities up to $0.1-0.2c$ \citep{Foltz:1983,Weymann:1991}. The BAL absorption troughs indicate the presence of energetic outflows in these systems, possibly launched from the accretion disk \citep{Murray:1995}. Powerful AGN outflows have been suggested to have a profound impact on their host galaxies by regulating and potentially quenching its star formation \citep{Silk:1998,Fabian:2012,Zubovas:2012}. However, observationally the role, ubiquity and impact of AGN outflows is still poorly understood. BAL QSOs may provide a useful probe to constrain the AGN outflow mechanism.

In optically selected samples the BAL fraction is $\sim15$\% \citep{Hewett:2003,Gibson:2009}, but their intrinsic fraction is likely higher \citep{Urrutia:2009,Allen:2011,Maddox:2012}. Narrower absorption features are also seen even more commonly in the UV-lines of quasars, \revt{which are classified based on their full-width-half-maximum (FWHM) as either narrow absorption lines (NAL, FWHM$<500$~km~s$^{-1}$) or mini-BALs \citep[$500<\mathrm{FWHM}<2000$~km~s$^{-1}$,][]{Hamann:1997,Vestergaard:2003,Misawa:2007}.}

%Narrower absorption features are also seen even more commonly in the UV-lines of quasars, which are classified as either narrow absorption lines (NAL) (\rev{full-width-half-maximum (FWHM)} $<500$~km~s$^{-1}$) or mini-BALs ($500<\mathrm{FWHM}<2000$~km~s$^{-1}$) \citep{Hamann:1997,Vestergaard:2003,Misawa:2007}.

\revt{The BALs are most commonly observed only  in high-ionization lines such as \ion{C}{4} and \ion{Si}{4} in which case the AGN are called \rev{high-ionization broad absorption line quasars (HiBALs)}. Less common ($\sim15$\%) are \rev{low-ionization broad absorption line quasars (LoBALs)}, which in addition also show absorption in low-ionization lines like \ion{Mg}{2}  and \ion{Al}{3}.} 
\rev{For both BAL QSO populations, both an orientation scenario and an evolution scenario have been proposed. HiBAL QSOs are commonly explained with an orientation scenario \citep{Weymann:1991,Ogle:1999,Gallagher:2007}, while a pure orientation effect is not sufficient to explain in particular observations for the radio-loud BAL QSO population \citep[e.g.][]{Bruni:2012, DiPompeo:2012, DiPompeo:2013}. For LoBALs more frequently an evolution scenario is advocated} \citep{Boroson:1992b,Voit:1993}, in which they constitute an early, short-lived transition phase of AGN activity, with observational support for \citep{Boroson:1992b,Canalizo:2002,Farrah:2007} and against this scenario \citep{Lazarova:2012,Violino:2016,Schulze:2017}. The evolution scenario has been most  strongly argued for the rare sub-population within the LoBAL class  of FeLoBALs,  which in addition also show absorption troughs in the metastable \ion{Fe}{2} line \citep{Hazard:1987,Becker:1997,Hall:2002}. 

\begin{deluxetable*}{lcccccccccccc}
\tabletypesize{\small}
%\rotate
\tablecaption{Multi-band photometry for the two LoBAL QSOs}
\tablewidth{18cm}
\tablehead{\colhead{Object} & \colhead{$u$} & \colhead{$g$} & \colhead{$r$} & \colhead{$i$} & \colhead{$z$} & \colhead{$J$} & \colhead{$H$} & \colhead{$K$} & \colhead{$W1$} & \colhead{$W2$} & \colhead{$W3$} & \colhead{$W4$}
}
\startdata
SDSS J1019+0225 & 21.02 & 20.49 & 19.58 & 18.48 & 18.00 & 16.37 & 15.22 & 15.11 & 14.06 & 12.82 &  9.61 &  6.95 \\
SDSS J0859+4239 & 19.91 & 19.24 & 19.33 & 18.38 & 18.02 & 16.49 & 15.38 & 15.85 & 14.73 & 13.54 & 10.68 &  7.50 \\
\enddata
\tablecomments{The $ugriz$ magnitudes are given in AB magnitudes in the SDSS system, the $JHK$ magnitudes are Vega-magnitudes in the 2MASS system and the $W1-W4$ magnitudes are Vega magnitudes in the WISE system.}
 \label{tab:photo}
\end{deluxetable*}

Currently, the rarest case of absorption features in BALs is absorption in their Balmer lines, with only 11 cases reported to date, \rev{ of which only two are known at $z>1.0$} \citep{Hutchings:2002,Hall:2002,Aoki:2006,Hall:2007,Wang:2008,Aoki:2010,Ji:2012,Ji:2013,Wang:2015,Zhang:2015,Mudd:2017}. %Only two of these are known at $z>1$ \citep{Aoki:2006,Aoki:2010}. 
The existence of Balmer absorption requires a high number density of $n=2$ level hydrogen atoms, which might be caused by Ly$\alpha$ trapping \citep{Hall:2007}. The physical conditions for this to happen are presumably rare, leading to the observed rarity of Balmer absorption lines in AGN.

We here present the discovery of two luminous LoBAL QSOs at $z\sim1.5$ which show strong Balmer absorption features in their near-IR spectra, namely SDSS J085910.40+423911.3 and SDSS J101927.37+022521.4, hereafter SDSS J0859+4239 and SDSS J1019+0225. Both objects have been observed as part of the sample presented in \citet{Schulze:2017} (hereafter S17). \revt{In S17 we obtained near-IR spectroscopy, covering H$\alpha$, H$\beta$ and H$\gamma$, for a sample of 12 LoBAL QSOs at $z\sim1.5$ and 10 at $z\sim2.3$. The targets were selected from the Sloan Digital Sky Survey (SDSS) quasar catalog \citep{Schneider:2010},  showing continuous broad absorption in  \ion{Mg}{2} or  \ion{Al}{3} as measured by \citet{Allen:2011} using the balnicity index \citep[BI;][]{Weymann:1991}.}
%In S17 we obtained near-IR spectroscopy, covering H$\alpha$, H$\beta$ and H$\gamma$, for a sample of 12 LoBAL QSOs at $z\sim1.5$ with non-zero \ion{Mg}{2} balnicity index and another 10 at $z\sim2.3$ with non-zero \ion{Al}{3} balnicity, drawn from the catalog of \citet{Allen:2011}, based on the Sloan Digital Sky Survey (SDSS) quasar catalog \citep{Schneider:2010}. \rev{The balnicity index (BI) is a measure of the strengths of a continuous broad absorption feature as defined by \citet{Weymann:1991} and is taken from \citet{Allen:2011}.}
We identified the two Balmer absorption systems within these objects.

Throughout this paper we assume a Hubble constant of $H_0 = 70$ km s$^{-1}$ Mpc$^{-1}$ and cosmological density parameters $\Omega_\mathrm{m} = 0.3$ and $\Omega_\Lambda = 0.7$.

\section{Observations}
Both SDSS J0859+4239 and SDSS J1019+0225 have been observed with the near-IR spectrograph TripleSpec \citep{Wilson:2004} at the Palomar Hale 200 inch telescope in January 2014 with an exposure time of 60~min each. TripleSpec provides simultaneous $JHK$ coverage from 1.0 $\mu$m to 2.4 $\mu$m at a spectral resolution of R$\sim2700$ and uses a slit width of 1\arcsec. The data reduction is carried out using the modified IDL-based Spextool3 package \citep{Cushing:2004}, and is described in more detail in S17 \citep[see also][]{Zuo:2015}.

Optical spectroscopy comes originally from SDSS DR7 \citep{Abazajian:2009}. For SDSS J0859+4239 \rev{we use} a more recent eBOSS spectrum from SDSS DR14 \citep{Abolfathi:2017}, which extends to longer wavelengths. This spectrum is consistent with the older SDSS spectrum, indicating no strong spectral variability over the time span of 14 years from Jan. 2002 to Dec. 2015. % (see also Fig.~\ref{fig:spec_uv}).

\rev{Furthermore, we} utilize information on the broad band spectral energy distribution (SED) from the UV to the mid-infrared, based on photometry in the optical $ugriz$ bands from SDSS \citep{Schneider:2010}, near-infrared in the $JHK$ bands from 2MASS \citep{Skrutskie:2006} and mid-IR from the all-sky Wide-Field Infrared Survey Explorer \rev{\citep[\textit{WISE}; ][]{Wright:2010,Lang:2016}} mission at $3.4, 4.6, 12$ and $22 \mu$m respectively. We provide the multi-band photometry for our two objects in Table~\ref{tab:photo}.

\begin{figure*}
{\centering
% \resizebox{\hsize}{!}{\includegraphics[clip]{baspec_j1019.eps} } }
\includegraphics[width=13cm,clip]{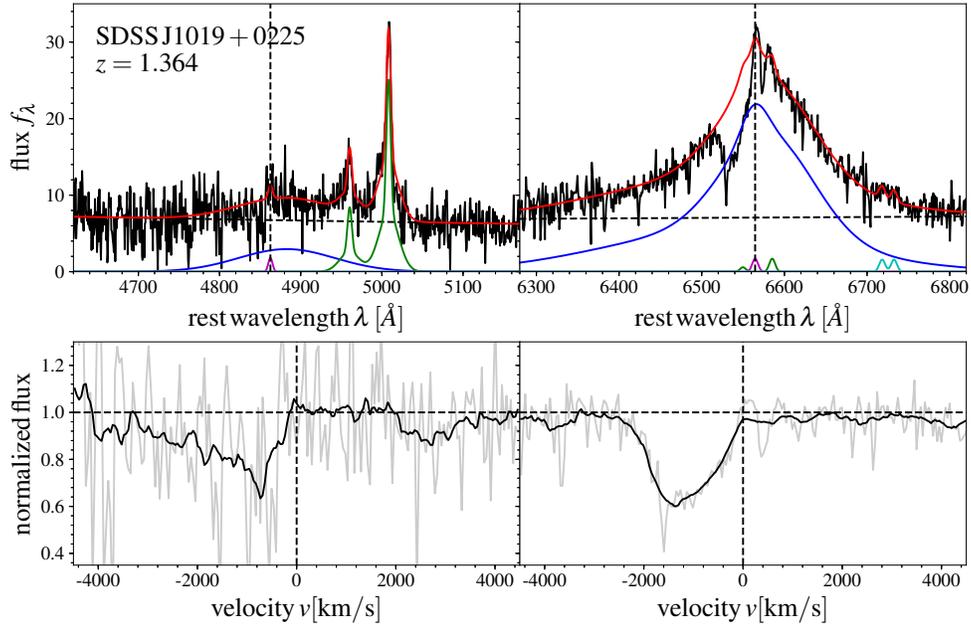}\\ }
\caption{Upper panels: Near-IR spectrum for SDSS J1019+0225, covering the H$\beta$ (left) and H$\alpha$ (right) region. We show our best fit continuum+emission line model (red solid line), where the Balmer absorption region has been masked out. The model includes a power-law continuum (black dashed line), a multi-Gauss model for the broad Balmer lines (blue) and [\ion{O}{3}] (green) as well as single Gaussian components for the narrow Balmer lines (magenta), [\ion{N}{2}] (green) and [\ion{S}{2}] (cyan). The flux scale is in units of $10^{-17}$~erg~cm$^{-2}$~s$^{-1}$~\AA$^{-1}$.
Lower panels: Normalized spectrum in velocity space, derived by dividing the observed flux by the best fit model, which highlights the Balmer absorption line trough. %\rev{The redshift zero-point the intrinsic [\ion{O}{3}] redshift is used.}
The gray line shows the original spectrum, while the black lines shows a smoothed version of it. The vertical dashed line indicates the velocity zero-point, while the horizontal dashed line gives \rev{the level of the continuum and the broad emission lines.}}
\label{fig:spec1}
\end{figure*}

\begin{figure*}
{\centering
% \resizebox{\hsize}{!}{\includegraphics[clip]{baspec_j1019.eps} }  }
\includegraphics[width=13cm,clip]{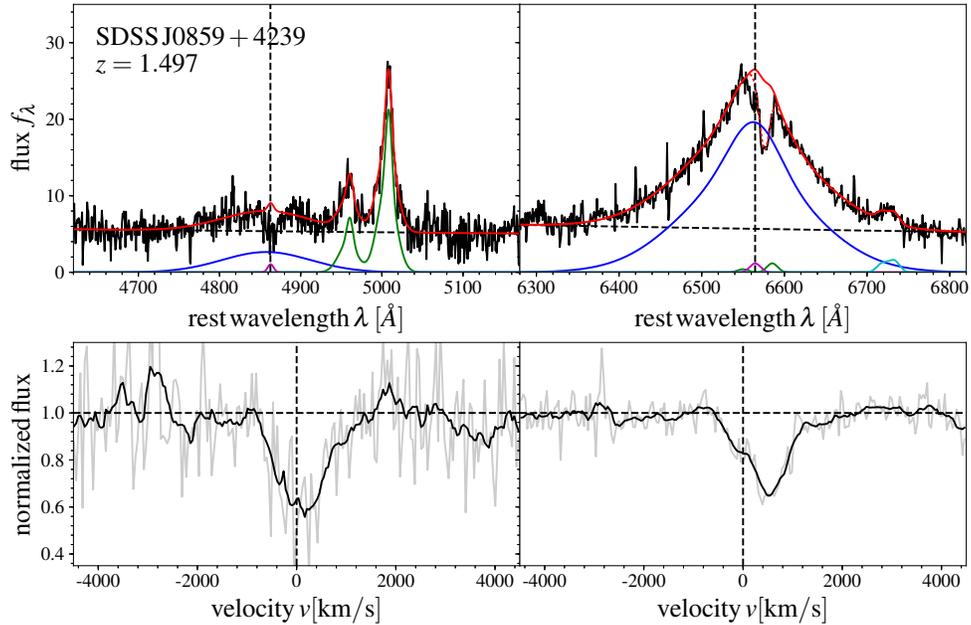} \\ }
\caption{Same as Fig.~\ref{fig:spec1}, but for SDSS J0859+4239. In addition we note that for H$\alpha$ we have included the absorption line into the fit as a single Gaussian component (red dashed line).}
\label{fig:spec2}
\end{figure*}

\section{Results}
The H$\alpha$ and H$\beta$ spectral regions obtained from the near-IR spectra are shown in the upper panels in Fig.~\ref{fig:spec1} for SDSS J1019+0225 and in Fig.~\ref{fig:spec2} for SDSS J0859+4239. Both objects show the clear presence of absorption troughs in their Balmer lines. We also see tentative evidence for Balmer absorption in H$\gamma$, though at low significance. % (see Fig.~\ref{fig:spec_abs}).

We have fitted the wavelength regions around H$\alpha$ and H$\beta$ with a multi-component spectral model, as discussed in more detail in S17. In short, we fit a local power-law continuum, an optical iron template \citep{Boroson:1992} for the H$\beta$ region and a set of Gaussian components for the broad Balmer lines, two Gaussian components for the [\ion{O}{3}] lines and single Gaussians for the narrow Balmer lines, [\ion{N}{2}] $\lambda\lambda6548,6584$ and [\ion{S}{2}] $\lambda\lambda6717,6731$. For SDSS J1019+0225 and for H$\beta$ in SDSS J0859+4239 we have masked out the spectral range of the Balmer absorption, while for H$\alpha$ in SDSS J0859+4239 we included it in our spectral model as a single Gaussian absorption line. 
The redshift of the objects is derived from the peak flux in the [\ion{O}{3}] $\lambda5007$ line.

In the lower panels of Fig.~\ref{fig:spec1} and Fig.~\ref{fig:spec2} we present the spectra normalized to this spectral model fit in velocity space as gray line, which shows the normalized absorption trough in respect to the broad line and continuum emission. %To further highlight the absorption lines, we smooth the normalized spectra by an uniform filter, shown as black solid line. 
%To further highlight the absorption feature, we show as black solid line the normalized spectra smoothed by a Gaussian filter, using a Gaussian kernel of $\sim4$\AA{} and $\sim2$\AA{} width for SDSS~J1019+0225 and SDSS~J0859+4239, respectively. 
To further highlight the absorption feature, we show as black solid line the normalized spectra smoothed by an uniform filter, computing the local mean over an uniform kernel of $\sim13$\AA{} and $\sim8$\AA{} rest-frame width for SDSS~J1019+0225 and SDSS~J0859+4239, respectively. 
We perform our measurements of the absorption trough based on this smoothed spectrum. We have also explored smoothing with a Gaussian filter and found  consistent results.

\rev{In particular, we used this smoothed spectrum} to measure the velocity $V_\mathrm{peak}$ and absorption depth $D_\mathrm{peak}$ of the peak position of the absorption trough. Furthermore, we measure \rev{FWHM} (corrected for instrumental broadening) the rest-frame equivalent width (EW$_\mathrm{rest}$) in respect to the continuum+broad emission line and the minimum and maximum velocity. The latter two are defined as the location in velocity space where the normalized flux falls below 95\%. The EW$_\mathrm{rest}$ is measured over the velocity region between the minimum and maximum velocity.
The absorption trough measurements are listed in Table~\ref{tab:meas}.

\rev{Uncertainties on the absorption trough measurements are derived via Monte Carlo simulations. For each spectrum we generate 100 simulated
spectra by adding Gaussian random noise to the spectra, with the standard deviation at each pixel taken from the flux error. For each simulated spectrum we carry out the same measurements as for the real spectrum, including model fitting, normalizing, smoothing and  absorption trough measurements. The uncertainties are then taken as the 1$\sigma$ dispersion from those measurments.}

We note that both objects have strong [\ion{O}{3}] emission, with EW$_\mathrm{rest}=54$\AA{} and 80\AA{} for  SDSS~J1019+0225 and SDSS~J0859+4239, respectively. \rev{LoBAL QSOs are traditionally thought to show very weak [\ion{O}{3}]  emission \citep{Boroson:1992b}, like the greater population of BAL QSOs \citep{Yuan:2003}. \revt{Conversely,}} Balmer absorption QSOs often have strong [\ion{O}{3}] emission lines \citep{Aoki:2006,Hall:2007}. However, in S17 we showed that at least our LoBAL QSO sample at $z\sim1.5$ shows a broad range of [\ion{O}{3}] equivalent widths, consistent with the general quasar population. \rev{Balmer absorption QSOs with weak  [\ion{O}{3}] have also} been found \citep{Ji:2013,Zhang:2015}. \rev{This suggests that normal LoBALs and Balmer absorption line LoBALs do not have clearly separated [\ion{O}{3}]  equivalent width distributions, as early observations suggested, but likely overlap in their [\ion{O}{3}] properties.}  Nevertheless we note that our two Balmer absorption line cases posses the strongest [\ion{O}{3}] emission among our sample in S17, which might indicate that LoBAL QSOs with Balmer absorption line systems  show on average higher  [\ion{O}{3}] equivalent width than the general LoBAL QSO population. Larger Balmer absorption line AGN samples are required to robustly test this trend.

\revt{Both objects also do not show prominent iron emission in their rest-frame optical, given the data quality of our near-IR spectra. This is consistent with the on average weak iron emission seen for the full LoBAL QSO sample in S17, but in contrast to the typically strong iron emission in LoBAL QSOs at $z<1$.}

We next discuss the two individual Balmer absorption line LoBAL QSOs in more detail.

\subsection{SDSS J1019+0225}
SDSS J1019+0225 is a LoBAL with \ion{Mg}{2} Balnicity of 4648~km~s$^{-1}$ \citep{Allen:2011}, located at redshift $z=1.364$ (S17, based on  [\ion{O}{3}]$\lambda5007$ and fully consistent with  [\ion{O}{2}]). It has a black hole mass and Eddington ratio of  $\log \mbh=9.87\ [M_\odot]$ and  $\log L_{\rm{bol}}/L_{\rm{Edd}}=-0.86$, as obtained in S17 from the broad H$\alpha$ line\footnote{\rev{Note that black hole mass estimates using the virial method as done here likely have a systematic uncertainty of $\sim0.3$~dex.}}. 

SDSS J1019+0225 is detected in the Faint Images of the Radio Sky at Twenty Centimeters \citep[FIRST;][]{Becker:1995} survey. \revt{It is originally classified as radio loud, with a radio-loudness parameter $R=f_{6\rm{cm}}/f_{2500}=129$ \citep{Shen:2011}. However, this value is affected by the significant reddening of the spectrum. To account for this we use the less affected flux at 5100\AA{} and estimate $f_{2500}$ by assuming a ratio $f_{2500}/f_{5100}=4.5$, based on the quasar template from  \citep{Shen:2016}. This gives $R_\mathrm{intrinsic}=10.8$, i.e. using the common definition of a radio loud QSO of $R>10$ \citep{Kellermann:1989}, SDSS J1019+0225 is just above the boundary to be classified as radio loud. We note that radio loud QSOs show on average stronger [\ion{O}{3}]  emission than matched radio quiet QSOs \citep[e.g.][]{Schulze:2017b}, consistent with the relatively strong  [\ion{O}{3}] emission observed in SDSS J1019+0225. On the other hand, small samples of radio loud BAL QSOs found typically  week  [\ion{O}{3}]  emission \citep[][]{Runnoe:2013}.
}

%\rev{Radio loud BAL QSOs are usually rare \citep[e.g.][]{Becker:2000,Bruni:2014}, so  a radio loud LoBAL QSO with Balmer absorption as SDSS J1019+0225 is  a very rare object. However, we note that SDSS J1019+0225 has a very massive black hole and in general the radio loud fraction significantly increases for such high black hole masses \citep{Laor:2000,Kratzer:2015}. Radio loud QSOs also show on average stronger [\ion{O}{3}]  emission than matched radio quiet QSOs \citep[e.g.][]{Schulze:2017b}, which might be a reason for the relatively strong [\ion{O}{3}]  emission observed in SDSS J1019+0225.}

\begin{deluxetable*}{lccccccc}
\tabletypesize{\small}
%\rotate
\tablecaption{Balmer Absorption line measurements}
\tablewidth{18cm}
\tablehead{
\colhead{Object} & \colhead{Line} & \colhead{EW$_\mathrm{rest}$}  & \colhead{FWHM} & \colhead{$V_\mathrm{peak}^a$} & \colhead{$D_\mathrm{peak}$} &  \colhead{$V_\mathrm{min}^a$} &  \colhead{$V_\mathrm{max}^a$} \\
\colhead{}            &  \colhead{}    & \colhead{[\AA{}]}  & \colhead{[km/s]} & \colhead{[km/s]} & \colhead{} & \colhead{[km/s]} & \colhead{[km/s]}
}
\startdata
SDSS J1019+0225 & H$\alpha$ & $12.6^{+0.6}_{-0.4}$  & $1457^{+42}_{-42}$ & $-1386^{+84}_{-42}$ & $0.60^{+0.01}_{-0.02}$ & $-2346^{+84}_{-125}$ & $-51^{+42}_{-42}$ \\
                               & H$\beta$  & $8.0^{+1.8}_{-4.6}$   & $1303^{+5506}_{-633}$ &  $-731^{+45}_{-2706}$  & $0.63^{+0.03}_{-0.14}$ & $-3212^{+1669}_{-857}$ & $-190^{+90}_{-42}$ \\
SDSS J0859+4239 & H$\alpha$ & $7.3^{+0.3}_{-0.2}$ & $783^{+199}_{-40}$ & $511^{+40}_{-1}$ & $0.65^{+0.01}_{-0.01}$ & $-438^{+1}_{-40}$ & $1222^{+40}_{-1}$ \\
                               & H$\beta$  & $7.9^{+0.9}_{-0.7}$  & $1061^{+3077}_{-129}$ &  $164^{+1}_{-256}$  & $0.56^{+0.03}_{-0.05}$ & $-732^{+85}_{-85}$ & $1231^{+128}_{-171}$ \\
\enddata                               
\tablecomments{ $a - $The velocity is defined as negative for lines that are blueshifted from the quasar. }
 \label{tab:meas}
\end{deluxetable*}

The presence of an asymmetric absorption profile is visible in both the broad H$\alpha$ and H$\beta$ emission line, however the S/N in H$\beta$ is rather poor. The trough in H$\alpha$ is blueshifted by $\sim1400$~km s$^{-1}$ and has an EW$_\mathrm{rest}\sim13$\AA{}. This is the strongest Balmer absorption feature currently known at $z>1$. \rev{The FWHM of the Balmer absorption trough} is $\sim1500$~km s$^{-1}$, making \rev{SDSS J1019+0225} a mini-BAL in its Balmer lines.

The Balmer absorption is intrinsic to the quasar and not due to a post-starburst feature in its host galaxy. Post-starburst features in quasars are not uncommon \citep{Brotherton:1999,Cales:2013} and have also recently been observed in a LoBAL QSO \citep{Mudd:2017}. \rev{However, the Balmer absorption in this system is clearly stronger that the stellar continuum emission  (which is subdominant to the total continuum emission in this luminous quasar) and thus cannot be stellar absorption. Furthermore, Balmer absorption in post-starburst quasars is usually not blueshifted in respect to the [\ion{O}{3}] and [\ion{O}{2}] emission}. Post-starburst quasars also show a strong Balmer break, which for SDSS J1019+0225 however falls into an observed wavelength range strongly affected by atmospheric absorption.

\subsection{SDSS J0859+4239} \label{sec:J0859}
SDSS J0859+4239 has a redshift of $z=1.497$ (also based on  [\ion{O}{3}] $\lambda5007$ and fully consistent with  [\ion{O}{2}]) and a  \ion{Mg}{2} Balnicity of 5212~km s$^{-1}$ \citep{Allen:2011}. Its black hole mass and Eddington ratio are $\log \mbh=10.11\ [M_\odot]$ and  $\log L_{\rm{bol}}/L_{\rm{Edd}}=-1.09$, respectively (S17). It is also detected in FIRST \revt{with $R_\mathrm{intrinsic}=2.4$ \citep[$R=15$ in][]{Shen:2011}, i.e. it is classified as radio quiet after correcting for dust reddening.} 

SDSS J0859+4239 shows a strong, rather unique Balmer absorption feature. The trough has a EW$_\mathrm{rest}\sim7$\AA{}, comparable to the Balmer absorption LoBAL discovered by \citet{Aoki:2006}. However, unlike almost all other Balmer absorption features the trough in SDSS J0859+4239 is redshifted instead of blueshifted, by $\sim500$~km s$^{-1}$ in respect to the [\ion{O}{3}] and [\ion{O}{2}] redshift. Most Balmer absorption systems are blueshifted up to 5000-10000~km s$^{-1}$  from its emission redshift \citep{Zhang:2015}, indicative of a fast outflowing wind. The only redshifted Balmer absorption system reported yet is SDSS~J112526.12+002901.3 \citep{Hall:2002,Shi:2016}, showing absorption lines redshifted up to 650~km s$^{-1}$ from the systemic redshift  in the Balmer lines and \ion{He}{1}$^*$. \citet{Shi:2016} proposed this absorption system as a candidate for accretion inflow which originates from the inner surface of the torus. Until now observational evidence for strong gas inflows in AGN on physical scales within the torus is scarce. If the Balmer absorption in SDSS J0859+4239 is indeed associated with a gas inflow onto the central black hole, which is seen along our line of sight, this would support the idea proposed by \citet{Shi:2016} and provide a unique case to study the fueling of supermassive black holes. Alternative explanations for or contributions to redshifted absorption systems are a rotation-dominated disk wind or gravitational redshift \citep{Hall:2002,Hall:2013}. Possible ways to discriminate between these scenarios are variability or constraining the location of the absorption system (see also section \ref{sec:loc}).

We classify the Balmer absorption system in SDSS J0859+4239 as a mini-BAL, given its FWHM of 783~km s$^{-1}$ in H$\alpha$. As for SDSS J1019+0225, we argue that the absorption feature is intrinsic and not indicative of a post-starburst feature for the same reasons as above. The absorption  in both H$\alpha$ and H$\beta$ is deeper than the height of the continuum level, which already excludes an origin in the AGN host galaxy.

\begin{figure}
{\centering
 \resizebox{\hsize}{!}{\includegraphics[clip]{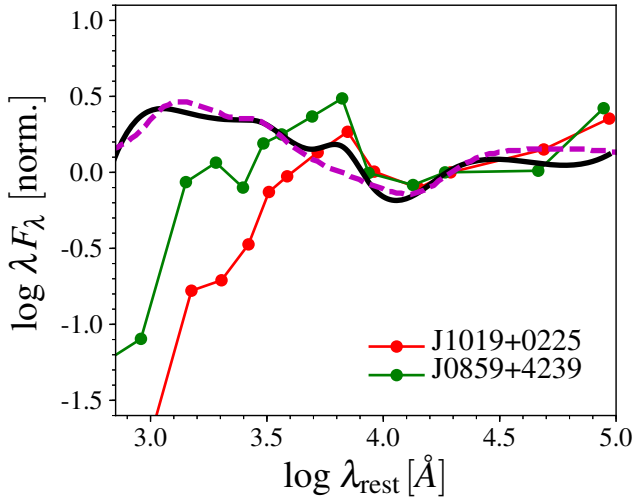} } }
\caption{Spectral energy distribution (SED) for J1019+0225 (red) and SDSS J0859+4239 (green), normalized at $K$-band. We compare these with the geometric mean SED for a matched non-BAL sample, presented in \citet{Schulze:2017} (black solid line) and the quasar SED by  \citet{Richards:2006} (purple dashed line).}
\label{fig:sed}
\end{figure}

\begin{figure*}
{\centering
% \resizebox{\hsize}{!}{\includegraphics[clip]{baspec_j1019.eps} }
\includegraphics[width=18cm,clip]{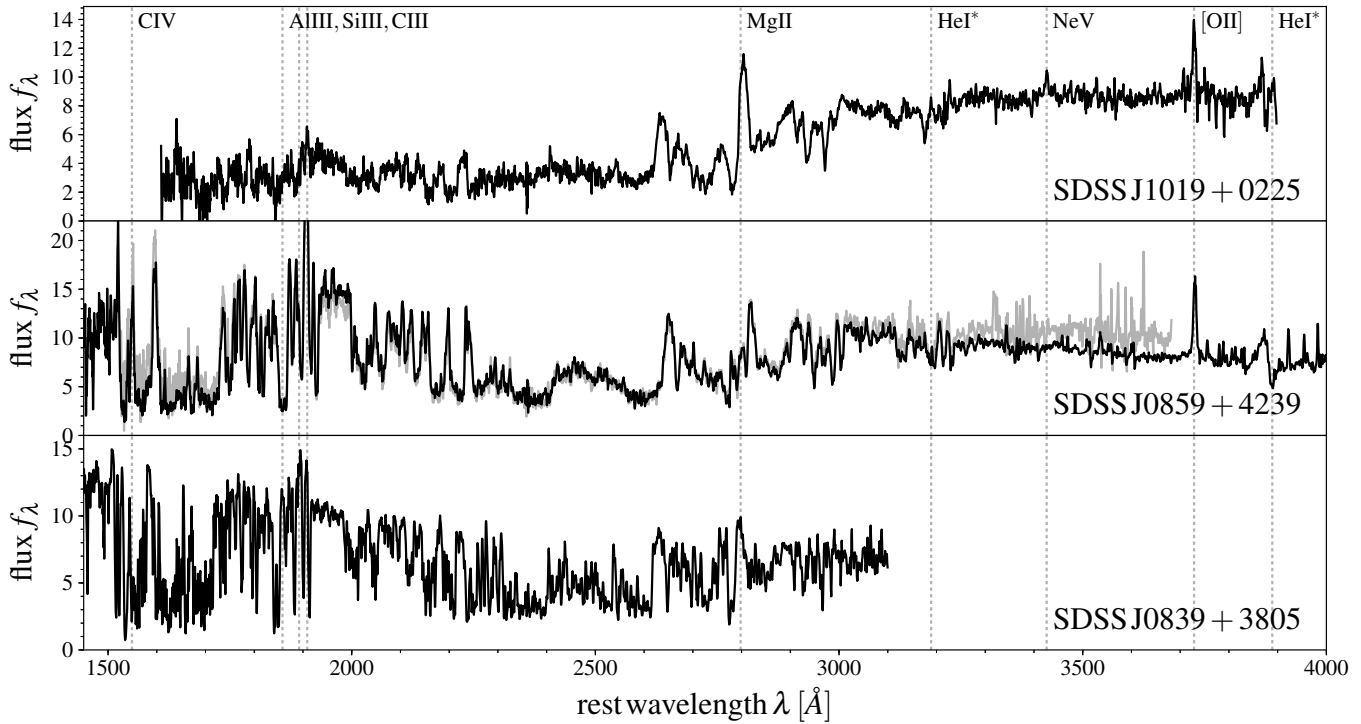} }
\caption{Rest-frame UV spectra from SDSS and eBOSS for our two Balmer absorption LoBALs and for SDSS~J0839+3805, a Balmer absorption FeLoBAL at $z=2.3$ discovered by \citet{Aoki:2006}. We indicate the location of several prominent lines based on the [\ion{O}{3}] redshift as vertical dotted lines. For SDSS~J0859+4239 we also show the original SDSS spectrum in gray below the more recent eBOSS spectrum.}
\label{fig:spec_uv}
\end{figure*}

\subsection{SED and Reddening}
LoBAL QSOs typically are strongly reddened, with a dust extinction of $E(B-V)\sim0.14$ \citep{Sprayberry:1992,Gibson:2009}. This is also the case for our two Balmer absorption LoBALs, as seen from their SED, shown in Fig.~\ref{fig:sed}. We construct the SED for the two objects, based on the multi-band photometry given in Table~\ref{tab:photo}, as explained in S17, and normalized them to $K$-band. In addition, we show the AGN SED template from \citet{Richards:2006} and the median SED for a sample of normal quasars matched to our LoBAL sample in S17 as the purple dashed line and black solid line respectively.

Similar to our results in S17 for the full LoBAL sample, we find the rest frame mid-IR to optical SED consistent with that of normal quasars, while the rest frame UV is strongly affected by reddening. Assuming  SMC-like dust extinction we find $E(B-V)\sim0.25$ and $\sim0.1$ for SDSS~J1019+0225 and SDSS~J0859+4239 respectively\rev{, by matching the un-reddened SED to the SED of normal quasars.}

In addition, we also estimate the dust reddening from the optical and near-IR spectrum, by correcting for SMC-like dust extinction until achieving an approximate match to an unreddened AGN  template. We use the quasar composite spectrum from \citet{Shen:2016}, which is derived for a comparable luminosity and redshift range. We also tested the SDSS quasar composite spectrum from \citet{VandenBerk:2001}, finding consistent results. We estimate dust extinction values from the spectra of $E(B-V)\sim0.33$ and $\sim0.1$ for SDSS~J1019+0225 and SDSS~J0859+4239 respectively. The former is larger than the value estimated from the full SED, while the latter is fully consistent with this estimate.

\section{Discussion}

\subsection{Absorption in the rest-frame UV lines}
Both objects show complex absorption systems from multiple lines in their rest-frame UV spectra, as shown in Fig.~\ref{fig:spec_uv}. We show the SDSS DR7 spectrum for SDSS~J1019+0225 and a more recent SDSS DR14 spectrum for SDSS~J0859+4239.
We have indicated the position of several prominent AGN emission lines, based on the [\ion{O}{3}] redshift. We identify complex absorption troughs from \ion{Mg}{2}, \ion{Al}{3} and \ion{He}{1}$^*$. Furthermore, we see strong [\ion{O}{2}] emission in both cases and a prominent \ion{Ne}{5} line in SDSS~J1019+0225. 

\begin{figure}
{\centering
\resizebox{\hsize}{!}{\includegraphics[clip]{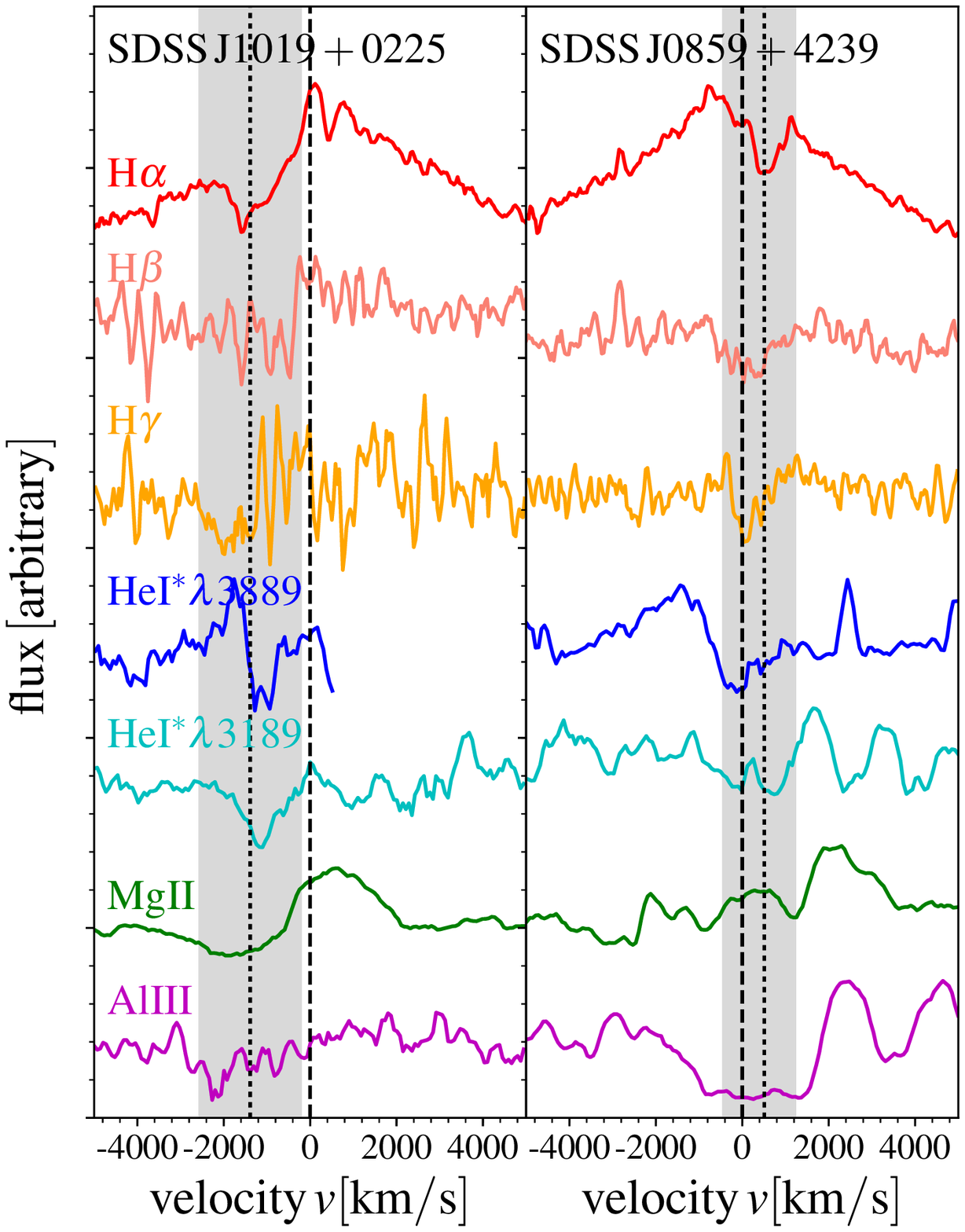} } }
\caption{Absorption troughs in several low-ionization lines in velocity space, including the Balmer lines (H$\alpha$, H$\beta$, H$\gamma$, \ion{He}{1}$^*$ at 3889 and 3189\AA{},  \ion{Mg}{2} and \ion{Al}{3}. The vertical dashed lines indicate the velocity zero-point, based on the [\ion{O}{3}] redshift. The dotted lines and gray area mark the peak velocity and the extend of the H$\alpha$ trough respectively. The spectra have been arbitrarily shifted in flux and rescaled in some cases for better visibility.
}
\label{fig:spec_abs}
\end{figure}

Absorption in the \ion{He}{1}$^*$ line has been thought to be a rare phenomenon \citep{Rudy:1985,Hall:2002,Leighly:2011}. However, the systematic study by \citet{Liu:2015} suggested associated \ion{He}{1}$^*$ absorption to be common among LoBAL QSOs, with an observed fraction $>90$\% for SDSS quasars with high S/N spectra. \ion{He}{1}$^*$ lines originate from the metastable triplet $2s$ level caused by recombination of \ion{He}{2}, with their strongest transitions at 10830, 3889, and 3189\AA{}. \ion{He}{1}$^*$ absorption can be a powerful tool to probe the physical conditions and geometry of the (outflowing) absorber \citep{Leighly:2011,Liu:2015}. It has been detected in the majority of the currently known Balmer absorption line AGN \citep{Zhang:2015}. We detect both \ion{He}{1}$^*\lambda3889$ and \ion{He}{1}$^*\lambda3189$ absorption troughs in the optical SDSS/eBOSS spectra for both objects.

In Fig.~\ref{fig:spec_abs} we compare the absorption features of several lines from rest-frame UV to optical with each other in velocity space, using the [\ion{O}{3}] line for the systemic redshift. The \ion{He}{1}$^*$ absorption at  both $\lambda3889$\AA{} and $\lambda3189$\AA{{} is in broad agreement with the H$\alpha$ absorption.
In SDSS~J0859+4239 we see tentative evidence for two absorption features, one blueshifted and one slightly redshifted, where the former one is weaker in H$\alpha$. However, in \ion{He}{1}$^*\lambda3189$ both appear to be equally strong, while in \ion{He}{1}$^*\lambda3889$ the blueshifted component is even stronger. Higher S/N observations would be required to robustly establish this line profile behavior. %{\bf Do we believe this? Do we have a physical interpretation for it?}

The troughs in  \ion{Mg}{2} and \ion{Al}{3} broadly agree as well in their velocity offset, but are much stronger and broader than the absorption in the Balmer and \ion{He}{1}$^*$ lines. Both objects have a high balnicity in \ion{Mg}{2}, BI$\sim5000$~km s$^{-1}$ \citep{Allen:2011}. The balnicity in \ion{Al}{3} for SDSS~J0859+4239 is $496$~km s$^{-1}$, while they did not detect an \ion{Al}{3} BAL in SDSS~J1019+0225, probably due to the low S/N in the SDSS spectrum.
Overall the velocity offset of the absorption troughs in the optical Balmer lines and the UV low-ionization lines are comparable, while we see indications for differences in their line profiles. Higher quality spectra would be required for a better characterization of in particular the \ion{He}{1}$^*$ line profile. A robust comparison of the line profiles of these absorption systems would provide better constraints on the structure (e.g possible stratification) and physical conditions of the outflowing/inflowing gas.
%The general agreement between the absorption in the optical Balmer lines and the UV low-ionization lines suggests that both are produced by the same outflowing/inflowing low-ionization gas. {\bf should we add more?}

In the bottom panel of Fig.~\ref{fig:spec_uv} we show in addition a BOSS spectrum (taken in 2010) for SDSS J083942.11+380526.3 (hereafter SDSS~J0839+3805), the first high-$z$ Balmer absorption quasar discovered \citep{Aoki:2006}. Interestingly, we find a close resemblance in the rest-frame UV spectra between SDSS~J0839+3805 and SDSS~J0859+4239. This might indicate similar physical conditions of the absorbing gas in both objects. SDSS~J0839+3805 does show H$\alpha$ absorption with a rest equivalent width of 8~\AA{}, comparable to SDSS~J0859+4239, and a blueshift of 520 km s$^{-1}$, while the absorption feature in SDSS~J0859+4239 is redshifted by about the same amount. Since SDSS~J0839+3805 is at $z=2.32$, its BOSS spectrum does not cover the \ion{He}{1}$^*$ lines, but \citet{Aoki:2006} report a tentative discovery of \ion{He}{1}$^*\lambda3889$ in their $J$-band spectrum, though at low significance.

\begin{deluxetable*}{lccccccc}
\tabletypesize{\small}
%\rotate
\tablecaption{Coverage fraction, optical depth and hydrogen column density estimates}
\tablewidth{18cm}
\tablehead{
\colhead{Object} & \colhead{Scenario} & \colhead{$C_f$} & \colhead{$\tau_{\rm{H}\alpha}$} & \colhead{$N_{\rm{H\,I}}(n=2)$} & \colhead{$\tau_{\rm{Ly}\alpha}$} & \colhead{$N_1/N_2$} &  \colhead{$N_{\rm{H\,I}}$} 
}
\startdata
SDSS J1019+0225 & I  & $0.40^{+0.23}_{-0.01}$  &  $6.7^{+1.3}_{-5.5}$ & $1.13^{+0.16}_{-0.48}\times10^{15}$ cm$^{-2}$ & $1199^{+126}_{-694}$ & $1490^{+2047}_{-175}$ & $1.69^{+0.74}_{-0.32}\times10^{18}$ cm$^{-2}$\\
                               &  II & $0.21^{+0.20}_{-0.07}$  &  $2.0^{+1.7}_{-1.3}$ & $3.58^{+1.05}_{-1.10}\times10^{14}$ cm$^{-2}$ & $661^{+235}_{-266}$ & $2702^{+1823}_{-709}$ & $9.68^{+2.57}_{-1.95}\times10^{17}$ cm$^{-2}$\\
SDSS J0859+4239 &  I & $0.34^{+0.01}_{-0.01}$  &  $5.3^{+0.2}_{-0.7}$ & $7.28^{+1.03}_{-2.64}\times10^{14}$ cm$^{-2}$ & $1067^{+24}_{-74}$ & $1675^{+125}_{-37}$ & $1.22^{+0.15}_{-0.43}\times10^{18}$  cm$^{-2}$ \\
                               &  II & $0.17^{+0.01}_{-0.02}$  &  $3.8^{+1.1}_{-0.5}$ & $2.94^{+0.29}_{-0.27}\times10^{14}$ cm$^{-2}$ & $901^{+122}_{-57}$ &  $1983^{+133}_{-236}$ & $5.82^{+0.29}_{-0.62}\times10^{17}$ cm$^{-2}$ \\                              
 \enddata
 \tablecomments{\rev{Derived physical properties of the absorber, as discussed in section~\ref{sec:nh}, using two different scenarios for the coverage of the absorber. The coverage fraction $C_f$ and H$\alpha$ optical depth $\tau_{\rm{H}\alpha}$ are derived based on equations~\ref{eq:cf1}-\ref{eq:r2}. The total column density of \ion{H}{1} gas in the $n=2$ level $N_{\rm{H\,I}}(n=2)$ is computed by integrating Equation~\ref{eq:dN}. We use this, together with the optical depth of Ly$\alpha$ and the ratio between the  $n=1$ to $n=2$ shell population $N_1/N_2$ (based on equations \ref{eq:tau_lya} and \ref{eq:n1n2}) to estimate the total hydrogen column density $N_{\rm{H\,I}}$.}}
 \label{tab:nh}
\end{deluxetable*}

\subsection{Hydrogen column density} \label{sec:nh}

Our near-IR spectroscopy of the Balmer absorption system allows to derive an estimate of the hydrogen column density of the absorber in both 
SDSS~J1019+0225 and SDSS~J0859+4239. A significant caveat here is the poor S/N detection of the H$\beta$ absorption. We use the smoothed absorption profiles shown in the bottom panels of Fig.~\ref{fig:spec1} and \ref{fig:spec2}, but  point out that higher quality spectra would be crucial for a more robust estimate.

Under the assumption of full coverage by the absorber along our line of sight to the background flux source the ratio of absorption depths at the centers of the H$\alpha$ and H$\beta$ lines should be close to the ratio of their oscillator strengths, $\sim5$ \citep{Zhang:2015}. However, our observations suggest a much smaller ratio, rather close to one.  \rev{The most likely explanation is that the absorption lines are} saturated and the covering factor, i.e. the fraction of flux covered by the foreground absorber, is smaller than one.\footnote{\rev{Possible alternative scenarios to explain such a deviation from the expected optical depth scaling have been discussed by \citet{Ganguly:1999}, but these are less likely.}}.
Thus the H$\alpha$ absorber does not fully cover the continuum source and the Broad emission line region (BELR). It might either cover only one of them fully, one fully and one partially, or both of them partially. Most likely the H$\alpha$ absorber covers at least the BELR partially because its absorption depth is larger than the continuum flux at that wavelength in both objects. An estimate of the covering factor can be derived from comparing the absorption depths of several lines with simple models \citep{Zhang:2015}.

We here follow this approach and specifically test two plausible scenarios for the partial coverage of the absorber of BELR and continuum disk emission: (I) the absorber covers the BELR and continuum source with the same covering factor; and (II) the absorber covers the continuum source fully (covering factor of one), but the BELR only partially. These two scenarios should approximately span the possible range of partial coverage for our sources.

For scenario (I), the effective coverage fraction $C_f$ can be computed from the normalized residual intensity $R$ in the trough, when observed in multiple lines \citep{Barlow:1997,Hamann:1997,Ganguly:1999}. The intensity is given by:
\begin{equation}
R(\lambda)= 1- C_f (\lambda)+ C_f(\lambda) e^{-\tau(\lambda)} \ ,  \label{eq:cf1}
\end{equation}
where $\tau$ is the effective optical depth of an absorbing cloud, covering a fraction $C_f$ of the background source. For the two transitions of H$\alpha$ and H$\beta$, $C_f$  can be derived via solving the relation:
\begin{equation}
\left[ \frac{R_\alpha - 1 + C_f}{C_f} \right]^{\frac{f_\beta \lambda_\beta}{f_\alpha \lambda_\alpha}}  =  \frac{R_\beta - 1 + C_f}{C_f} \ ,   \label{eq:r1}
\end{equation}
where $\lambda_\alpha$ and $\lambda_\beta$ are the rest-wavelengths of H$\alpha$ and H$\beta$ and $f_\alpha=0.640$ and $f_\beta=0.119$ are their oscillator strengths. \rev{The optical depth $\tau$ is then determined via Equation~\ref{eq:cf1}.}

For scenario (II), where we assume different coverage fractions of BELR and continuum source, equation~\ref{eq:cf1} is modified to  \citep{Ganguly:1999}
\begin{equation}
R= 1- \frac{(C_c + W C_f) (1-e^{-\tau})}{1+W} \ ,  \label{eq:cf2}
\end{equation}
where $C_c=1$ is the coverage fraction of the continuum source, $C_f$ in this case would be the coverage fraction towards the BELR and $W=F_{\rm{BEL}}/F_c$ is the ratio of the broad emission line flux to the continuum flux, each derived from our best fit model. For multiple transitions the following equation has to be solved \rev{to determine $C_f$}:
\begin{equation}
\left[ \frac{(R_\alpha - 1)(1+W_\alpha)}{C_c +W_\alpha C_f} +1 \right]^{\frac{f_\beta \lambda_\beta}{f_\alpha \lambda_\alpha}} =  \frac{(R_\beta - 1)(1+W_\beta)}{C_c +W_\beta C_f} +1\ .  \label{eq:r2}
\end{equation}
\rev{Again, $\tau$ is then computed via Equation~\ref{eq:cf2}.}
We determine coverage fraction $C_f$ and H$\alpha$ optical depth $\tau_{\rm{H}\alpha}$ for both scenarios from above equations for every velocity element from the smoothed absorption troughs. Their values at the velocity of the absorption peak are listed in Table~\ref{tab:nh} for our two objects.

The hydrogen column density in the $n=2$ level as a function of velocity is given by \citep{Arav:2001,Zhang:2015}
\begin{equation}
\frac{\mathrm{d}N_{\rm{H\,I}}}{\mathrm{d}v}=\frac{m_e c}{\pi e^2} \frac{1}{\lambda_\alpha f_\alpha} \tau_{\rm{H}\alpha}(v) \ . \label{eq:dN}
\end{equation}
We calculate the total column density of \ion{H}{1} gas in the $n=2$ level in the absorbing cloud by integrating Equation~\ref{eq:dN}. We obtain values of $N_{\rm{H\,I}}(n=2)=(3-11)\times 10^{14}$~cm$^{-2}$ for SDSS~J1019+0225 and $(3-7)\times 10^{14}$~cm$^{-2}$  for SDSS~J0859+4239 for the two scenarios (I) and (II), as listed in Table~\ref{tab:nh}.

We furthermore estimate the total neutral column density following \citet{Hall:2007} and \citet{Aoki:2010}. The optical depth of Ly$\alpha$ is related to those of H$\alpha$ via
\begin{equation}
\tau_{\rm{Ly}\alpha} = \frac{\lambda_{\rm{Ly}\alpha} f_{\rm{Ly}\alpha}}{\lambda_{\rm{H}\alpha} f_{\rm{H}\alpha}} \frac{N_1}{N_2} \tau_{\rm{H}\alpha} \ ,  \label{eq:tau_lya}
\end{equation}
where $f_{\rm{Ly}\alpha}=0.416$ is the oscillator strength of Ly$\alpha$ and $N_1$ and $N_2$ are the population of levels $n=1$ and $n = 2$, respectively. Due to Ly$\alpha$ trapping the $n=2$ level population is increased by a factor $\tau_{\rm{Ly}\alpha}$ from the thermal equilibrium. The ratio of the $n=1$ to $n=2$ shell is then \citep{Hall:2007}
\begin{equation}
\frac{N_1}{N_2} =\frac{1}{4} \exp(10.2 \mathrm{eV}/ kT) \tau_{\rm{Ly}\alpha}^{-1} \ .  \label{eq:n1n2}
\end{equation}
We assume $T=7500$~K, as appropriate for partially ionized gas illuminated by a quasar \citep{Osterbrock:2006}, to derive $\tau_{\rm{Ly}\alpha}$, $N_1/N_2$ and the total hydrogen column density $N_{\rm{H\,I}}$ for our two objects and both scenarios. The results are given in Table~\ref{tab:nh}. We find $N_{\rm{H\,I}}=(9-17)\times 10^{17}$~cm$^{-2}$ for SDSS~J1019+0225 and $(6-12)\times 10^{17}$~cm$^{-2}$  for SDSS~J0859+4239, \rev{using our results from scenario~(I) and (II) to approximately bracket the possible range of  $N_{\rm{H\,I}}$.}

\subsection{Locating the origin of Balmer absorption lines} \label{sec:loc}
Balmer absorption lines originate in low or partially-ionized  and high-density regions. The low ionization ensures a large amount of neutral hydrogen to be present. For high-density regions  \rev{($\rev{\log(n_e/}$ cm$\rev{^{-3}) \sim9}$)} then the optical depth for Ly$\alpha$ is so large that Ly$\alpha$ pumping \citep{Ferland:1979} is important, which keeps a large number of excited level ($n>2$) hydrogen atoms \citep{Hall:2007}. This indicates the absorber \rev{is located} very close to the central engine, typically on scales between the dust torus and the BELR \citep{Hall:2007,Zhang:2015}. \citet{Hutchings:2002} observed NGC4151 (a nearby Sy1 galaxy) and proposed that Balmer and \ion{He}{1}$^*$ absorption   lines originate around the edge of the obscuring torus that is eroded   and accelerated by the nuclear flux. \citet{Zhang:2015} and \citet{Shi:2016} also studied the Balmer  absorbers of low-$z$ quasars at $z < 1$ and confirmed their distances  from the central engine are larger than the size of the BELR but smaller than the size of the dusty torus based on detailed photoionization models using CLOUDY \citep{Ferland:1998}. \rev{These results suggest that outflow winds with the right physical conditions to produce Balmer absorption are most likely located just outside of the BELR. %These winds will have low ionization conditions and high density \rev{($\rev{\log(n_e/}$ cm$\rev{^{-3}) \sim9}$)}.
}
%These results suggest outflow winds with high density \rev{($\rev{\log(n_e/}$ cm$\rev{^{-3}) \sim9}$)} in low ionization condition, required to produce Balmer absorption, survive in the inner region of the AGN, just outside of the BELR. 
However, it has not been confirmed yet whether this geometry is also applicable to high-$z$ quasars. 

The two $z\sim1.5$ Balmer absorption LoBAL quasars presented here offer the potential to address this question directly. A detailed characterization of the multiple absorption lines detected in both the optical and near-IR in combination with careful photoionization modeling using CLOUDY is able to provide robust estimates of the hydrogen column density, optical depth and covering factor as well as the electron density, ionization parameter and location of the absorber. However, both the quality of the optical and the near-IR spectra presented here are, while being sufficient for detection, not sufficient for such a detailed modeling. Future, high S/N and high resolution spectroscopy will be required to derive these quantities.

\section{Conclusions}
We here present the discovery of two new cases of Balmer absorption lines in quasars from near-IR spectroscopy with Triplespec/Palomar. 
Balmer absorption troughs are currently rarely seen in quasar spectra. \rev{This suggests that the necessary physical conditions for such absorption to be present do rarely occur. This might be due to an orientation effect, where only a limited range of lines of sight allow the observation of Balmer absorption\revt{, but we are unable to rule out other explanations.}}
The two new cases of Balmer absorption lines, SDSS~J1019+0225 and SDSS~J0859+4239, are both luminous LoBAL QSOs at $z\sim1.5$, doubling the number of known Balmer absorption line quasars at $z>1$. We detect absorption in H$\alpha$, H$\beta$, tentatively in H$\gamma$, as well as in \ion{He}{1}$^*$ at $\lambda3889$ and $\lambda3189$, in addition to BAL absorption in 
\ion{Mg}{2} and \ion{Al}{3}. The Balmer absorption in SDSS~J0859+4239 is redshifted by $~\sim 500$~km s$^{-1}$, \rev{with} respect to the [\ion{O}{3}] redshift, potentially indicating an inflow of high density gas onto the black hole. We estimated the neutral hydrogen column densities, $N_{\rm{H\,I}}\sim 1.3\times 10^{18}$~cm$^{-2}$ for SDSS~J1019+0225 and $\sim9\times 10^{17}$~cm$^{-2}$  for SDSS~J0859+4239.

While from the current data we are not able to constrain the location of the absorber, it is expected that it is located on small scales, roughly between the BELR and the dusty torus \citep{Hall:2007,Zhang:2015}. Higher S/N and higher resolution data will be required to better constrain the physical conditions of the outflowing/inflowing absorber and their location for our two sources \rev{via photoionization modeling.}
%%%%%%%%%%%%%%%%%%%%%%%%%%%%%%%%%%%%%%%

%\begin{figure*}
%\centering
%\includegraphics[height=7cm,clip]{baspec_j1019.eps} \hspace{0.2cm}
%\includegraphics[height=7cm,clip]{spec_absorb_J1019.eps} \\
%\includegraphics[width=16.5cm,clip]{baspec_uv_J1019.eps} 
%\caption{ }
%\label{fig:eso}
%\end{figure*}

%%%%%%%%%%%%%%%%%%%%%%%%%%%%%%%%%%%%%%%

\acknowledgments
A.S. is supported by the EACOA fellowship and acknowledges support by JSPS KAKENHI Grant Number 26800098. This research was partially supported by the Japan Society for the Promotion of Science through Grant-in-Aid for Scientific Research 15K05020. X.-B.Wu thanks the supports by the NSFC grants No.11373008 and 11533001, the National Key Basic Research Program of China 2014CB845700, and from the Ministry of Science and Technology of China under grant 2016YFA0400703.

This research uses data obtained through the Telescope Access Program (TAP), which has been funded by the National Astronomical Observatories of China, the Chinese Academy of Sciences (the Strategic Priority Research Program "The Emergence of Cosmological Structures" Grant No. XDB09000000), and the Special Fund for Astronomy from the Ministry of Finance. 
Observations obtained with the Hale Telescope at Palomar Observatory were obtained as part of an agreement between the National Astronomical Observatories, Chinese Academy of Sciences, and the California Institute of Technology.
%\software{Spextool3 \citep{Cushing:2004}, NumPy, SciPy, AstroPy, Matplotlib}. %, TOPCAT \citep{Taylor:2005}}

%\appendix 
%\section*{Case of single paragraph}
%
%\section{Case of two or paragraphs}
%
%\section{Case of two or paragraphs}

%%%
% See the manual for the detail.
%%%


\begin{thebibliography}{}
% Journals(e.g. A\&A,ApJ,AJ,NMRAS,PASP ...)
% Authors, Year, Journal, Vol#, Page#
% Journal Title Abbreviation >> http://www.asj.or.jp/pasj/Jabb.html
\bibitem[Abazajian et al.(2009)]{Abazajian:2009} Abazajian, K.~N., Adelman-McCarthy, J.~K., Ag{\"u}eros, M.~A., et al.\ 2009, \apjs, 182, 543-558 
\bibitem[Abolfathi et al.(2017)]{Abolfathi:2017} Abolfathi, B., Aguado, D.~S., Aguilar, G., et al.\ 2017, arXiv:1707.09322 
\bibitem[Allen et al.(2011)]{Allen:2011} Allen, J.~T., Hewett, P.~C., Maddox, N., Richards, G.~T., \& Belokurov, V.\ 2011, \mnras, 410, 860
\bibitem[Aoki et al.(2006)]{Aoki:2006} Aoki, K., Iwata, I., Ohta, K., et al.\ 2006, \apj, 651, 84
\bibitem[Aoki(2010)]{Aoki:2010} Aoki, K.\ 2010, \pasj, 62, 1333 
\bibitem[Arav et al.(2001)]{Arav:2001} Arav, N., de Kool, M., Korista, K.~T., et al.\ 2001, \apj, 561, 118 
\bibitem[Barlow \& Sargent(1997)]{Barlow:1997} Barlow, T.~A., \& Sargent, W.~L.~W.\ 1997, \aj, 113, 136 
\bibitem[Becker et al.(1995)]{Becker:1995} Becker, R.~H., White, R.~L., \& Helfand, D.~J.\ 1995, \apj, 450, 559 
\bibitem[Becker et al.(1997)]{Becker:1997} Becker, R.~H., Gregg, M.~D., Hook, I.~M., et al.\ 1997, \apjl, 479, L93
\bibitem[Becker et al.(2000)]{Becker:2000} Becker, R.~H., White, R.~L., Gregg, M.~D., et al.\ 2000, \apj, 538, 72  
\bibitem[Boroson \& Green(1992)]{Boroson:1992} Boroson, T.~A., \& Green, R.~F.\ 1992, \apjs, 80, 109 
\bibitem[Boroson \& Meyers(1992)]{Boroson:1992b} Boroson, T.~A., \& Meyers, K.~A.\ 1992, \apj, 397, 442 
\bibitem[Brotherton et al.(1999)]{Brotherton:1999} Brotherton, M.~S., van Breugel, W., Stanford, S.~A., et al.\ 1999, \apjl, 520, L87
\bibitem[Bruni et al.(2012)]{Bruni:2012} Bruni, G., Mack, K.-H., Salerno, E., et al.\ 2012, \aap, 542, A13
\bibitem[Bruni et al.(2014)]{Bruni:2014} Bruni, G., Gonz{\'a}lez-Serrano, J.~I., Pedani, M., et al.\ 2014, \aap, 569, A87 
\bibitem[Cales et al.(2013)]{Cales:2013} Cales, S.~L., Brotherton, M.~S., Shang, Z., et al.\ 2013, \apj, 762, 90 
%\bibitem[Canalizo \& Stockton(2001)]{Canalizo:2001} Canalizo, G., \& Stockton, A.\ 2001, \apj, 555, 719 
\bibitem[Canalizo \& Stockton(2002)]{Canalizo:2002} Canalizo, G., \& Stockton, A.\ 2002, Mass Outflow in Active Galactic Nuclei: New Perspectives, 255, 195
\bibitem[Cushing et al.(2004)]{Cushing:2004} Cushing, M.~C., Vacca, W.~D., \& Rayner, J.~T.\ 2004, \pasp, 116, 362 
\bibitem[DiPompeo et al.(2012)]{DiPompeo:2012} DiPompeo, M.~A., Brotherton, M.~S., \& De Breuck, C.\ 2012, \apj, 752, 6 
\bibitem[DiPompeo et al.(2013)]{DiPompeo:2013} DiPompeo, M.~A., Brotherton, M.~S., \& De Breuck, C.\ 2013, \mnras, 428, 1565 
\bibitem[Fabian(2012)]{Fabian:2012} Fabian, A.~C.\ 2012, \araa, 50, 455 
\bibitem[Farrah et al.(2007)]{Farrah:2007} Farrah, D., Lacy, M., Priddey, R., Borys, C., \& Afonso, J.\ 2007, \apjl, 662, L59 
\bibitem[Ferland \& Netzer(1979)]{Ferland:1979} Ferland, G., \& Netzer, H.\ 1979, \apj, 229, 274 
\bibitem[Ferland et al.(1998)]{Ferland:1998} Ferland, G.~J., Korista, K.~T., Verner, D.~A., et al.\ 1998, \pasp, 110, 761 
\bibitem[Foltz et al.(1983)]{Foltz:1983} Foltz, C., Wilkes, B., Weymann, R., \& Turnshek, D.\ 1983, \pasp, 95, 341
\bibitem[Gallagher et al.(2007)]{Gallagher:2007} Gallagher, S.~C., Hines, D.~C., Blaylock, M., et al.\ 2007, \apj, 665, 157
\bibitem[Ganguly et al.(1999)]{Ganguly:1999} Ganguly, R., Eracleous, M., Charlton, J.~C., \& Churchill, C.~W.\ 1999, \aj, 117, 2594
\bibitem[Gibson et al.(2009)]{Gibson:2009} Gibson, R.~R., Jiang, L., Brandt, W.~N., et al.\ 2009, \apj, 692, 758 
\bibitem[Hall et al.(2002)]{Hall:2002} Hall, P.~B., Anderson, S.~F., Strauss, M.~A., et al.\ 2002, \apjs, 141, 267 
\bibitem[Hall(2007)]{Hall:2007} Hall, P.~B.\ 2007, \aj, 133, 1271 
\bibitem[Hall et al.(2013)]{Hall:2013} Hall, P.~B., Brandt, W.~N., Petitjean, P., et al.\ 2013, \mnras, 434, 222 
\bibitem[Hamann et al.(1997)]{Hamann:1997} Hamann, F., Barlow, T.~A., Junkkarinen, V., \& Burbidge, E.~M.\ 1997, \apj, 478, 80 
\bibitem[Hazard et al.(1987)]{Hazard:1987} Hazard, C., McMahon, R.~G., Webb, J.~K., \& Morton, D.~C.\ 1987, \apj, 323, 263
\bibitem[Hewett \& Foltz(2003)]{Hewett:2003} Hewett, P.~C., \& Foltz, C.~B.\ 2003, \aj, 125, 1784 
\bibitem[Hutchings et al.(2002)]{Hutchings:2002} Hutchings, J.~B., Crenshaw, D.~M., Kraemer, S.~B., et al.\ 2002, \aj, 124, 2543
\bibitem[Ji et al.(2012)]{Ji:2012} Ji, T., Wang, T.-G., Zhou, H.-Y., \& Wang, H.-Y.\ 2012, Research in Astronomy and Astrophysics, 12, 369 
\bibitem[Ji et al.(2013)]{Ji:2013} Ji, T., Zhou, H.-y., Wang, T.-g., \& Wang, H.-y.\ 2013, Chinese Astron. Astrophys., 37, 17
\bibitem[Kellermann et al.(1989)]{Kellermann:1989} Kellermann, K.~I., Sramek, R., Schmidt, M., Shaffer, D.~B., \& Green, R.\ 1989, \aj, 98, 1195
%\bibitem[Kratzer \& Richards(2015)]{Kratzer:2015} Kratzer, R.~M., \& Richards, G.~T.\ 2015, \aj, 149, 61
\bibitem[Lang et al.(2016)]{Lang:2016} Lang, D., Hogg, D.~W., \& Schlegel, D.~J.\ 2016, \aj, 151, 36 
%\bibitem[Laor(2000)]{Laor:2000} Laor, A.\ 2000, \apjl, 543, L111
\bibitem[Lazarova et al.(2012)]{Lazarova:2012} Lazarova, M.~S., Canalizo, G., Lacy, M., \& Sajina, A.\ 2012, \apj, 755, 29
\bibitem[Leighly et al.(2011)]{Leighly:2011} Leighly, K.~M., Dietrich, M., \& Barber, S.\ 2011, \apj, 728, 94 
\bibitem[Liu et al.(2015)]{Liu:2015} Liu, W.-J., Zhou, H., Ji, T., et al.\ 2015, \apjs, 217, 11 
\bibitem[Maddox et al.(2012)]{Maddox:2012} Maddox, N., Hewett, P.~C., P{\'e}roux, C., Nestor, D.~B., \& Wisotzki, L.\ 2012, \mnras, 424, 2876
\bibitem[Misawa et al.(2007)]{Misawa:2007} Misawa, T., Eracleous, M., Charlton, J.~C., \& Kashikawa, N.\ 2007, \apj, 660, 152 
\bibitem[Mudd et al.(2017)]{Mudd:2017} Mudd, D., Martini, P., Tie, S.~S., et al.\ 2017, \mnras, 468, 3682 
\bibitem[Murray et al.(1995)]{Murray:1995} Murray, N., Chiang, J., Grossman, S.~A., \& Voit, G.~M.\ 1995, \apj, 451, 498 
\bibitem[Ogle et al.(1999)]{Ogle:1999} Ogle, P.~M., Cohen, M.~H., Miller, J.~S., et al.\ 1999, \apjs, 125, 1
\bibitem[Osterbrock \& Ferland(2006)]{Osterbrock:2006} Osterbrock, D.~E., \& Ferland, G.~J.\ 2006, Astrophysics of gaseous nebulae and active galactic nuclei, 2nd.~ed.~by D.E.~Osterbrock and G.J.~Ferland.~Sausalito, CA: University Science Books, 2006, 
\bibitem[Richards et al.(2006)]{Richards:2006} Richards, G.~T., Lacy, M., Storrie-Lombardi, L.~J., et al.\ 2006, \apjs, 166, 470 
\bibitem[Rudy et al.(1985)]{Rudy:1985} Rudy, R.~J., Stocke, J.~T., \& Foltz, C.~B.\ 1985, \apj, 288, 531
\bibitem[Runnoe et al.(2013)]{Runnoe:2013} Runnoe, J.~C., Ganguly, R., Brotherton, M.~S., \& DiPompeo, M.~A.\ 2013, \mnras, 433, 1778
\bibitem[Schneider et al.(2010)]{Schneider:2010} Schneider, D.~P., Richards, G.~T., Hall, P.~B., et al.\ 2010, \aj, 139, 2360
\bibitem[Schulze et al. (2017)]{Schulze:2017} Schulze, A., Schramm, M., Zuo, W., et al.\ 2017, \apj, 848, 104 
\bibitem[Schulze et al.(2017b)]{Schulze:2017b} Schulze, A., Done, C., Lu, Y., Zhang, F., \& Inoue, Y.\ 2017, \apj, 849, 4 
\bibitem[Shen et al.(2011)]{Shen:2011} Shen, Y., Richards, G.~T., Strauss, M.~A., et al.\ 2011, \apjs, 194, 45
\bibitem[Shen(2016)]{Shen:2016} Shen, Y.\ 2016, \apj, 817, 55
\bibitem[Shi et al.(2016)]{Shi:2016} Shi, X.-H., Jiang, P., Wang, H.-Y., et al.\ 2016, \apj, 829, 96
\bibitem[Silk \& Rees(1998)]{Silk:1998} Silk, J., \& Rees, M.~J.\ 1998, \aap, 331, L1 
\bibitem[Skrutskie et al.(2006)]{Skrutskie:2006} Skrutskie, M.~F., Cutri, R.~M., Stiening, R., et al.\ 2006, \aj, 131, 1163 
%\bibitem[Spitzer(1978)]{Spitzer:1978} Spitzer, L.\ 1978, Physical processes in the interstellar medium, by Lyman Spitzer.~ New York Wiley-Interscience, 1978.~333 p.,  
\bibitem[Sprayberry \& Foltz(1992)]{Sprayberry:1992} Sprayberry, D., \& Foltz, C.~B.\ 1992, \apj, 390, 39
%\bibitem[Taylor(2005)]{Taylor:2005} Taylor, M.~B.\ 2005, Astronomical Data Analysis Software and Systems XIV, 347, 29 
\bibitem[Urrutia et al.(2009)]{Urrutia:2009} Urrutia, T., Becker, R.~H., White, R.~L., et al.\ 2009, \apj, 698, 1095 
\bibitem[Vanden Berk et al.(2001)]{VandenBerk:2001} Vanden Berk, D.~E., Richards, G.~T., Bauer, A., et al.\ 2001, \aj, 122, 549 
\bibitem[Vestergaard(2003)]{Vestergaard:2003} Vestergaard, M.\ 2003, \apj, 599, 116 
\bibitem[Violino et al.(2016)]{Violino:2016} Violino, G., Coppin, K.~E.~K., Stevens, J.~A., et al.\ 2016, \mnras, 457, 1371
\bibitem[Voit et al.(1993)]{Voit:1993} Voit, G.~M., Weymann, R.~J., \& Korista, K.~T.\ 1993, \apj, 413, 95 
\bibitem[Wang et al.(2008)]{Wang:2008} Wang, T., Dai, H., \& Zhou, H.\ 2008, \apj, 674, 668-675
\bibitem[Wang \& Xu(2015)]{Wang:2015} Wang, J., \& Xu, D.~W.\ 2015, \aap, 573, A15 
\bibitem[Weymann et al.(1991)]{Weymann:1991} Weymann, R.~J., Morris, S.~L., Foltz, C.~B., \& Hewett, P.~C.\ 1991, \apj, 373, 23 
\bibitem[Wilson et al.(2004)]{Wilson:2004} Wilson, J.~C., Henderson, C.~P., Herter, T.~L., et al.\ 2004, \procspie, 5492, 1295
\bibitem[Wright et al.(2010)]{Wright:2010} Wright, E.~L., Eisenhardt, P.~R.~M., Mainzer, A.~K., et al.\ 2010, \aj, 140, 1868-1881
\bibitem[Yuan \& Wills(2003)]{Yuan:2003} Yuan, M.~J., \& Wills, B.~J.\ 2003, \apjl, 593, L11
\bibitem[Zhang et al.(2015)]{Zhang:2015} Zhang, S., Zhou, H., Shi, X., et al.\ 2015, \apj, 815, 113 
\bibitem[Zubovas \& King(2012)]{Zubovas:2012} Zubovas, K., \& King, A.\ 2012, \apjl, 745, L34 
\bibitem[Zuo et al.(2015)]{Zuo:2015} Zuo, W., Wu, X.-B., Fan, X., et al.\ 2015, \apj, 799, 189 

\end{thebibliography}
\end{document}